\documentclass[superscriptaddress,aps,letterpaper,10pt,twocolumn,floats,showpacs,amsmath,amsfonts,amssymb,pre,nofootinbib]{revtex4-2}

\usepackage[utf8]{inputenc}

\usepackage{url,psfrag,graphicx}
\usepackage{dcolumn}
\usepackage{amsmath,amssymb}
\usepackage{bm}
\usepackage{hyperref}
\usepackage{float,epsfig,color}
\usepackage{times}

\usepackage{xcolor}

\begin{document}

\author{Jesús. M. Marcos}
\affiliation{Departamento de F\'{\i}sica, Universidad de Extremadura, 06006 Badajoz, Spain}
\author{Yifan Li}
\affiliation{The Wolfson Faculty Department of Chemical Engineering, Technion — Israel Institute of Technology, Haifa 3200003, Israel }
\author{Mark Fasano}
\affiliation{Department of Mathematical Sciences, New Jersey Institute of Technology, Newark, New Jersey 07102, USA}
\author{Javier A. Diez}
\affiliation{Instituto de F\'{\i}sica Arroyo Seco, Universidad Nacional del Centro de la Provincia de Buenos Aires, and CIFICEN-CONICET-CICPBA, Pinto 399, 7000, Tandil, Argentina}
\author{Linda J. Cummings}
\affiliation{Department of Mathematical Sciences, New Jersey Institute of Technology, Newark, New Jersey 07102, USA}
\author{Ofer Manor}
\affiliation{The Wolfson Faculty Department of Chemical Engineering, Technion — Israel Institute of Technology, Haifa 3200003, Israel }
\author{Lou Kondic}
\affiliation{Department of Mathematical Sciences, New Jersey Institute of Technology, Newark, New Jersey 07102, USA}
\email{kondic@njit.edu}
\title[]
  { Monte-Carlo based model for the extraction of oil from oil-water mixtures using wetting and surface acoustic waves}

\begin{abstract}
This work presents a Monte Carlo (MC) based microscopic model for simulating the extraction of oil from oil-in-water emulsions under the influence of surface acoustic waves (SAWs). The proposed model is a two-dimensional Ising-lattice gas model that employs Kawasaki dynamics to mimic the interactions between oil, water, and air, as well as external forces such as gravity and acoustic stress. By incorporating both acoustic streaming and acoustic radiation pressure, the model captures key experimental observations, including selective oil extraction and droplet motion under SAW excitation. The results highlight the critical role of acoustic radiation pressure in enabling oil film formation and detachment, governed by the balance between capillary and acoustic stresses. The study provides qualitative agreement with experimental findings and offers insights into the essential mechanisms driving acoustowetting-induced phase separation, demonstrating the utility of discrete modeling for complex fluid dynamics problems.
\end{abstract}

\maketitle

\section{Introduction}

Common commercial techniques for oil-water separation, such as high-power distillation~\cite{Distillation} and coagulation/flocculation of oil droplets~\cite{ISI:16,ISI:59} (methods that have been in use for nearly two centuries), require large quantities of energy per product, as well as additional chemicals. Recently, it was shown that surface effects can be useful for oil-water separation, with important implications for energetic requirements: membranes of specific affinity for water or oil support the passage of the favorable phase through the membrane pores, while repelling the other phase through capillary effects~\cite{zhang2012smart,chen2013mineral}. 
For example, a hydrophobic poly vinylidene fluoride (PVDF) membrane was used to separate various water-in-oil emulsions including surfactant-free and surfactant-stabilized emulsions with droplet sizes from the micron to the nanometer range~\cite{zhang2013superhydrophobic}, and a hydrophilic hydrogel polyacrylamide was coated by a mesh consisting of rough nanostructured hydrogel coatings and microscale porous metal substrates to separate various oil-in-water mixtures~\cite{xue2011novel}. In addition, Zhang et al. \cite{zhang2012smart} have presented membranes of variable oil wettability based on polyurethane sponges, and Wen et al. \cite{wen2013zeolite} presented a zeolite-coated mesh film for oil-water separation. 
 

Oil is characterized by low surface tension, e.g., commercial silicon oil at ambient conditions supports a surface tension of approximately 20 mN/m at an interface with air. Water, on the other hand, is typically associated with higher surface tension: pure water at ambient conditions supports a surface tension of approximately 70 mN/m. Adding surfactants lowers the surface tension of the water/surfactant mixture; e.g., adding sodium dodecyl sulfate---used abundantly for commercial and domestic applications---to water reduces its surface tension down to approximately 40 mN/m at the critical micelle concentration.  Consequently, oil supports only a small three-phase (vapor/liquid/solid) contact angle with most solids; silicon oil in particular has vanishing contact angle at equilibrium with most solids, leading to many applications due to its ability to spontaneously spread over and coat large surface areas. Water and water/surfactant solutions support finite three-phase contact angles with most substrates, leading to the formation of drops~\cite{mittal_guide_2009}.

Recent experimental work extended the idea of harnessing surface contributions to enhance oil-water separation by introducing acoustic stress in the oil-water mixture, thereby generating a capillary/acoustic stress balance favorable to displacing oil from the mixture~\cite{exp}. 
One such mechanism of lately increasing popularity in the scientific literature is acoustic streaming: an acoustic wave (or ultrasonic wave---we use the term `acoustic' to refer to both wave regimes) traveling in a fluid, or in a solid neighboring a fluid, introduces stress and flow. It generates a boundary layer flow near the solid/fluid interfaces~\citep{LordRayleigh1884, Schlichting:1932p447} and a bulk flow in the fluid~\citep{Eckart:1948to,LIGHTHILL:1978p12,Nyborg:2004p312}. The bulk flow, whose steady component at long times is also known as Eckart streaming, is a product of intensity variations in the acoustic wave in the fluid,
which may attenuate due to viscous and thermal dissipation. Spatial variations in the wave intensity may also result from spatial variations in the source of the acoustic wave, as in the case where surface acoustic waves (SAWs), which travel and attenuate along the solid substrate, diffract (leak) acoustic waves to the fluid; the intensity of the acoustic wave traveling in the fluid, before thermal and acoustic attenuation, originates from the local intensity of the SAW in the solid, where the acoustic wave was leaked. Spatial variations in an acoustic wave traveling through fluid generate spatial variations in the convective Reynolds stress---the forcing mechanism for flow. Eckart streaming is abundantly used for the actuation of fluids in micro-channels \citep{yeoannurev,Wixforth:2004p910,Wixforth:2003jp} and drop microfluidics \citep{Guttenberg:2004wg,Brunet07,Brunet09,brunet_droplet_2010}. Moreover, the interaction of the acoustic waves with an interface, in our case the vapor/liquid interface of liquid drops and films, yields a net force on the interface---an acoustic radiation pressure~\cite{hamilton1998nonlinear}. This is known to introduce stress at the surface of particles~\citep{King:1934tp} and other solids \citep{Chu,Anonymous:5ILrl7tg,Hasegawa:2000vy,PhysRevLett.117.114504,Subramani2023} and was shown to deform and displace soft interfaces \citep{Biwersi:2000ti,Alzuaga05,issenmann_bistability_2006, rajendran_theory_2022}.  

It has been shown that MHz-level SAWs, which travel along a solid substrate, actuate the dynamic wetting of oil~\cite{Rezk2012,Rezk:2014,Manor} and water~\cite{altshuler2015spreading,altshuler2016free} films along and opposite the path of the SAW. The acoustic stress interacts differently with water and oil due to their different surface tensions and the corresponding capillary pressure these liquids experience. When the acoustic stress in the film dominates the capillary stress, liquid films dynamically wet a solid along and opposite the path of a SAW. This is trivial to achieve in the case of silicon oil due to its low surface tension. For water and water-surfactant mixtures to do the same, the SAW intensity must be above a certain threshold. Horesh et al.~\cite{Horesh} compared the response of oil and water-surfactant solutions to SAW excitation by incorporating gravitational effects into the acoustic and capillary stress balance. Oil films continuously climbed a vertical SAW actuator against gravity, whereas water and water-surfactant solutions only ascended a few millimeters before reaching an equilibrium height determined by the interplay of gravity, capillary stress, and acoustic stress.

Recently, \citet{exp} extended the previous work to study the extraction of oil films off oil-in-water mixtures (emulsions) in the laboratory. The oil phase was found to leave the mixture in the direction opposite the SAW propagation, since the acoustic stress therein overwhelmed the capillary stress associated with the low oil surface tension. The water phase remained at rest, since the capillary stress associated with the higher surface tension liquid was greater than the acoustic stress. Thus, oil films left the emulsion, while the water phase was left behind.

Simulating the dynamics of such oil-water separation under SAW excitation can provide deeper insight into the separation process. In this work, we propose a simple Monte Carlo (MC)-based microscopic model to better understand the key factors driving the experimentally observed phenomena. Various MC-based methods have already been implemented to simulate droplet behavior, with our focus specifically on discrete models where interactions are represented through a Hamiltonian that accounts for all relevant interactions in the system. Such discrete models have been analyzed using MC simulations under particle number conservation (Kawasaki dynamics) to capture particle exchange dynamics \cite{DeConinck1993, Cheng1993, Lukkarinen1995, Abraham2002, Marcos2022, Nussbaumer2008, Chalmers2017, Areshi2019}. Given the numerous possible interactions that define fluid dynamics, the system's interactions and geometry vary for each specific problem, depending on the phenomena being analyzed. The general approach taken in the cited works (among many others)  is to simplify the problem as much as possible, while retaining essential features. Below, we briefly review the studies most relevant to the problem considered here.

In the context of spreading, several decades ago~\citet{DeConinck1993} introduced a discrete two-dimensional Ising lattice gas model that included nearest-neighbor interaction between particles and an interaction with the substrate to investigate how a macroscopic droplet spreads. In the same year, \citet{Cheng1993} employed a slightly modified model, differing in the treatment of the substrate interaction, to examine the growth rate of a spreading droplet across a wide range of parameter values. A few years later, \citet{Lukkarinen1995} introduced a three-dimensional Ising model to investigate droplet spreading upon contact with a planar substrate, focusing on the emergence of an ultrathin precursor film. This model also incorporates nearest-neighbor interactions within an external field generated by the substrate potential and is defined on a cubic lattice of infinite extent in the in-plane directions, with finite extent in the out-of-plane direction. 

Other authors have employed similar models to describe various droplet behaviors beyond spreading. For example, the droplet formation-dissolution transition has been studied by \citet{Nussbaumer2008} using a simple two-dimensional Ising-lattice gas model, which only accounts for nearest-neighbor interactions. Later, \citet{Chalmers2017} investigated the evaporation of nanoparticle suspensions using a similar discrete model. This model incorporates pair interactions that extend beyond the nearest neighbors for the three possible interactions in the system: liquid-liquid, nanoparticle-nanoparticle, and liquid-nanoparticle. The model also incorporates a chemical potential governing the vapor-liquid phase transition, and distinct substrate interactions for nanoparticles and liquid particles. 
More recently, Areshi {\it et al.}~\cite{Areshi2019} used a similar model that includes pair interactions beyond the nearest neighbors, as well as substrate interaction, to study several aspects of droplet dynamics on surfaces. Specifically, they studied the density profiles, the mechanism by which two drops merge, and the evolution of droplets under a constant lateral driving force parallel to the surface. 

In all these models, the spin characteristics of the Ising model are replaced by occupancy numbers, where a value of zero indicates an empty lattice cell and unity signifies an occupied cell. Additionally, these models examine the system in a lattice setting, either 2D or 3D, where each cell can be occupied by only one particle at a time; we will follow a similar approach in this work.   

The rest of this paper is organized as follows. We proceed by presenting our model, with an emphasis on modeling the interactions between the different particle types considered. Our main goal is to explore the simplest approach to describing separation of oil and water from an emulsion using a discrete model similar to those discussed above, with a focus on considering emulsion drops exposed to external forcing by an applied surface acoustic wave (SAW). The computational results obtained from the MC simulations are then discussed in the Results section, which is followed by our Conclusions. Some technical details are given in the Appendices.

\section{Model}\label{Model}

While there have been previous attempts to model acoustowetting experiments using continuum theory, e.g.~\citet{altshuler2015spreading,altshuler2016free}, these works shed little light on the processes by which the SAW enhances the phase separation of an oil--in--water emulsion and do not describe the dynamics of either phase separation or oil film extraction. Therefore, to obtain new insight into these issues, we here take an alternative approach, using a simplified discrete two-dimensional model to describe the observed phenomena. The model that we simulate represents an idealized version of an oil--in--water emulsion and is based on energy minimization of a closed system of interacting, discrete, oil and water regions (cells within the model, described in what follows), where SAW-induced stress in the liquid (emulsion) is represented as an external force. Despite the model's simplicity, we will show that it is effective in qualitatively describing the recently reported experiments~\cite{exp} and providing valuable physical insights.

Our model is motivated by a system comprising a sessile drop of an oil--in--water emulsion positioned on top of a solid horizontal substrate, exposed to a SAW. The model consists of a two-dimensional regular lattice defining cells, which may be occupied or not, as discussed below.  The evolution of the system is simulated by a Monte Carlo scheme with Kawasaki local dynamics \cite{Newman1999}, which attempts to decrease the system energy by swapping the content of the neighboring cells. 
The main features of this established algorithm are described in the Appendix~\ref{ap:details}.  

The geometry of our model comprises a rectangular grid of $L_{x}$ cells along the $x-$axis (parallel to the solid substrate) and $L_{y}$ cells along the $y-$axis (normal to the solid substrate). The model can be easily generalized to three dimensions, however, in this work we restrict ourselves to two dimensions for simplicity. All cells are of the same size, and each is occupied by water, oil, or neither (modeling air). The basic setup is similar to that used by~\citet{Areshi2019}. As a point of clarification, we note that this description of the fluid is statistical rather than fully atomistic. A particle should be interpreted as a collective representation of multiple fluid molecules rather than a single molecule. The presence or absence of a particle at a lattice site reflects an increased or decreased probability of local occurrence of fluid molecules there. For further discussion, see~\citet{Chalmers2017} and references therein. The advantage of this modeling approach lies in its ability to efficiently analyze the system's interfacial properties while accurately incorporating the essential structure and thermodynamics of the fluid, despite its simplified nature \cite{Areshi2019}.

We define $o_{\bm{i}}$ and $w_{\bm{i}}$ as the occupation numbers of oil and water, respectively, so, for example, $o_{\bm{i}} = 1$ if the cell $\bm{i}$ is occupied by an oil particle and $o_{\bm{i}} =0 $ if it is occupied by water, or not occupied at all.  Here, $\bm{i}=(x_{\bm{i}},y_{\bm{i}})$ is the 2D position vector. 
We do not define an occupation number for air, as we will consider that air does not interact with other particles and is unaffected by the governing forces; an air cell is one for which $o_{\bm{i}} = 0 = w_{\bm i}$.

\begin{figure*}[!t]
\centering
\includegraphics[width=0.9\textwidth]{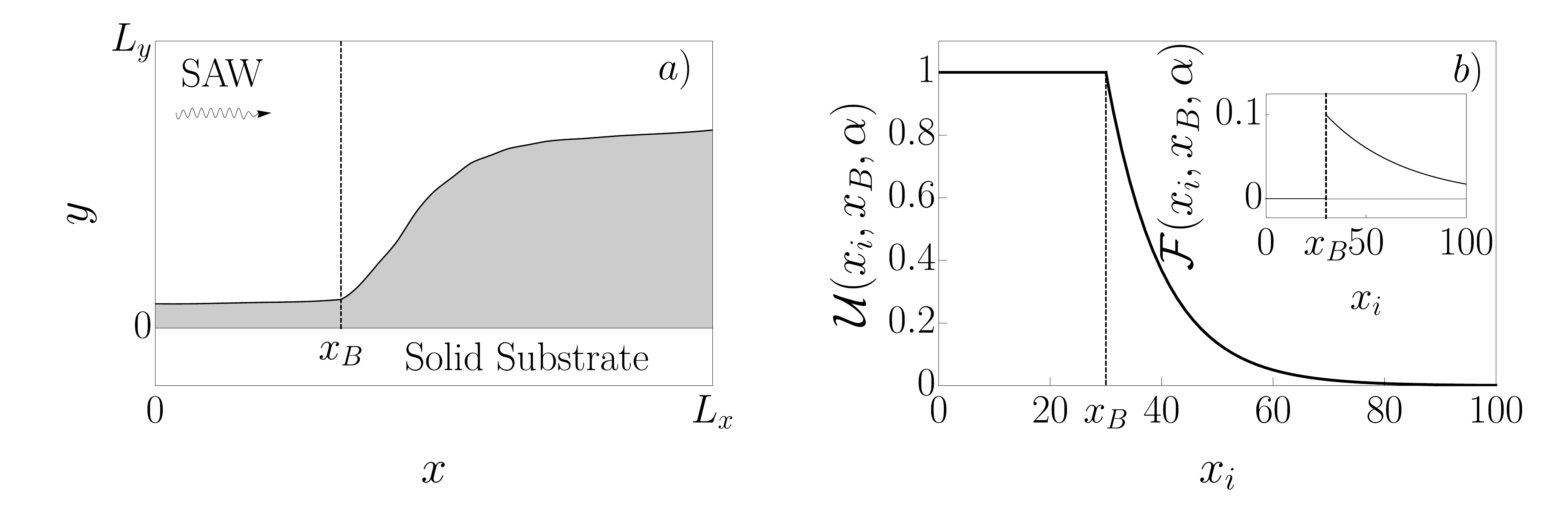}
\caption{(a) A sketch of the liquid (oil and water) geometry studied, where $x_B$ indicates the transition between an oil film ($x<x_B$) and the emulsion drop ($x>x_B$).  We also indicate the direction of SAW propagation. (b) Spatial variation of the acoustic potential in the liquid, given by $\mathcal{U}(x_{\bm{i}},x_B,\alpha)$ in Eq.~\eqref{eq:potential_energy_1}, where the dashed line indicates the position of $x_B$. Inset: Spatial variation of the force  $\mathcal{F}=\partial\mathcal{U}/\partial x$ generated in the liquid by the acoustic stress.}
\label{fig:extForce}
\end{figure*}

Initially, we assume a sessile drop, comprising a random mixture of water and oil particles, where we set the initial oil volume fraction to $c= \sum_i o_{\bm i}/ \sum_i (o_{\bm i} + w_{\bm i}) = 0.4$, as in the experiments~\cite{exp}. The energy of the system is obtained by summing over the contributions from all cells and accounts for close-neighbor interactions, external forcing due to gravity, and acoustic stress due to the SAW in the solid substrate. Each cell is assumed to have four nearest neighbors in the $x$ and $y$ directions, and another four next-nearest neighbors in the diagonal directions; water-water, oil-oil and oil-water interactions are accounted for. Although one could simply consider only nearest neighbor interactions, the motivation for including next-nearest neighbors comes from prior works  \cite{Areshi2019,Chalmers2017}, which noted that the use of only nearest neighbors can lead to the formation of unrealistic rectangular-shaped droplets at low temperatures. The resulting energy is given by the Hamiltonian 
\begin{equation}
    \begin{aligned}
        \mathcal{H} = & -\sum_{\langle \bm{i}, \bm{j} \rangle}c_{\bm{i}\bm{j}}\left( J_{oo} o_{\bm{i}}o_{\bm{j}} + J_{ww} w_{\bm{i}}w_{\bm{j}} + J_{ow} o_{\bm{i}}w_{\bm{j}}\right) \\
        & + \sum_{\bm{i}} \rho_{\bm{i}} g y_{\bm{i}} +\sum_{\bm{i}}\rho_{\bm{i}} p_{RS} \,\mathcal{U}(x_{\bm{i}},x_{B},\alpha).
    \end{aligned}
    \label{eq:energy}
\end{equation}
Here, the right hand side (RHS) includes the contributions due to the interaction energy between the cells, the gravitational contribution, and the acoustic term, respectively.  We describe each term below, first noting that, for simplicity, we consider each term on the RHS to be given in units of $k_B T$, where $k_B$ is the Boltzmann constant and $T$ absolute temperature.  To further simplify, we put $\rho_{\bm{i}} = 1$ (for both oil and water, as discussed below), $\rho_{\bm{i}} = 0$ for air, and we assign a numerical value to $g$ that is convenient for the purpose of our simulations (corresponding simply to a specific choice of scalings for the variables).  Clearly, more careful treatment of the relative magnitude of various physical effects will be needed if one attempts to model the quantitative details of any particular physical experiment.  

In Eq.~(\ref{eq:energy}), the first term on the RHS  describes the interaction energy between cells of index $\bm{i}$ and their neighbors of index $\bm{j}$, wherein ($\sum_{\langle\bm{i} \bm{j}\rangle}$) denotes summation over all pairs of cells on the grid. This term captures the interaction of water and oil cells that interact via a near-neighbor model via attractive forces and is reminiscent of van der Waals interactions between water and oil.  Here, the interaction energy with air is neglected, as we assume that the low (zero in our analysis) density air does not contribute to the total interaction energy. More precisely, interactions between water cells, between oil cells, and between water and oil cells, are described by energy contributions $J_{ww},~J_{oo}$ and $J_{wo}\equiv J_{ow}$, respectively. A larger (smaller) value of any of these positive parameters represents a stronger (weaker) interaction between these cells leading to a stronger (weaker) cohesive force. The interaction energy between pairs of particles at lattice sites $\bm{i}$ and $\bm{j}$ depends on the distance between them and is given by 
$c_{\bm{i}\bm{j}}$, which we define as \cite{Areshi2019}
\begin{equation}
    c_{\bm{i}\bm{j}}=\left\{
\begin{array}{ll}
      1 & \text{if $\bm{j}\in$ NN$\bm{i}$}\,, \\
      1/2 & \text{if $\bm{j}\in$ NNN$\bm{i}$} \,,\\
      0 & \text{otherwise}\, , \\
\end{array} \right.
    \label{eq:cij}
\end{equation}
where NN$\bm{i}$ and NNN$\bm{i}$ denote the nearest neighbors and the next nearest neighbors of the lattice site $\bm{i}$, respectively. Due to the negative sign before this term, the system will tend to evolve to a state that maximizes the number of interactions (or bonds) associated with the largest $J_{kl}$ ($k,l=o,w$). Since cells on the surface of the droplet can bond to fewer cells, the species with the highest interaction energies will tend to bond in the interior of the drop, while those with the lowest interaction energies will migrate to the surface. Therefore, by changing the parameters determining the near--neighbor interactions (discussed in what follows), we alter the effective interfacial energy, i.e., the surface tension, between oil and water. 

The second summation in Eq.~\eqref{eq:energy} represents the gravitational contributions to the potential energy of each cell $\bm{i}$, where $\rho_{\bm{i}}$ and $1\le y_{\bm{i}}\le L_y$ are the density and the (discrete) $y$-coordinate of the cell $\bm{i}$, respectively. The density of air is taken to be zero, and the densities of oil and water are taken to be the same (since our focus here is on the acoustic forcing, we ignore the buoyancy that results from the slight density difference between oil and water).

The third summation is the novel component of our model and accounts for the acoustic stress, i.e., the Reynolds stress in the liquid due to the SAW, which propagates in the positive $x$-direction along the substrate. The acoustic stress is assumed to be proportional to the density of the fluid cell (hence is negligible in air) and to the SAW power in the solid substrate.  It contains a factor $p_{RS}$ representing the acoustic stress in a cell that contains water or oil (assuming that both have similar acoustic impedance) due to an unattenuated SAW in the underlying solid; and a factor $\mathcal{U}$ that accounts for the attenuation of the SAW in the regions under the droplet. 
It is known, see e.g.~\citet{exp}, that a SAW decays exponentially under a sufficiently thick layer of fluid. We model this decay by specifying the appropriate dependence of $\mathcal{U}$ on $x_{\bm{i}}$, the (discrete) $x$-coordinate along the solid surface. Thus, in this simplified analysis, we ignore the Rayleigh angle at which ultrasonic waves leak off the SAW into the liquid and simply relate the acoustic stress in the liquid to the intensity of the SAW in the solid immediately below.  This approach is found to be appropriate for films whose thickness is small compared to the wavelength of the acoustic wave leakage off the SAW in the solid~\cite{Rezk2012,Rezk:2014,Manor}. It is somewhat naive for the macro-scale drop considered by~\citet{exp} where, in reality, the acoustic leakage off the SAW travels at the Rayleigh angle in the liquid body and supports a more complex stress field; however, we expect that a simplified treatment is sufficient to accommodate our main goals. Therefore, as a first approximation, we assume that the potential associated with the acoustic stress in liquid (oil or water) is of the form
\begin{equation}
    \mathcal{U}(x_{\bm{i}},x_B,\alpha) =\left\{\begin{matrix}
1 & x_{\bm{i}}<x_B ,\\ 
e^{-\alpha(x_{\bm{i}}-x_B)} & x_{\bm{i}}>x_B,
\end{matrix}\right.
    \label{eq:potential_energy_1}
\end{equation}
where $x_B$ is the coordinate at the edge of the drop, see Fig.~\ref{fig:extForce}, and $\alpha$ is the attenuation coefficient associated with the SAW. Below, we discuss further the key features of the SAW potential, to arrive at the final form used in our simulations.

Recalling that SAW attenuation becomes appreciable when the SAW travels under the macroscopic drop but is negligible elsewhere~\cite{Rezk2012,Rezk:2014}, in our simulations we define the macroscopic drop to be present at the $x_{\bm{i}}$ locations such that the fluid layer is consistently more than two cells thick. Therefore, the drop edge, $x_B$, is defined as the value of $x_{\bm{i}}$ at which the film thickness transitions from two cells to three and remains at least three until reaching the main body of the droplet. To compute this position, in each step of the algorithm, we loop through the system from left to right until we find a film thickness consistently greater than two. The exact choice for the number of liquid cells that define the threshold thickness is not important as long as it is small compared to the maximum drop thickness; choosing other suitably small values to define $x_B$ leads to similar results.  Thus, we ignore the SAW attenuation in $x_{\bm{i}}<x_B$, setting $\mathcal{U}=1$ in this region, and assume that attenuation occurs only for $x_{\bm{i}}>x_B$, where the SAW travels under the macroscopic drop. Here, we set $\mathcal{U}=e^{-\alpha(x_{\bm{i}}-x_B)}$, modeling the exponential attenuation of the acoustic stress in the liquid body, where $1/\alpha$ represents the characteristic attenuation length of the SAW. Figure~\ref{fig:extForce} shows $\mathcal{U}$ and the corresponding force factor by which the SAW acts on the liquid, $\mathcal{F}=-{\partial\mathcal{U}}/{\partial x}$, as defined by Eq.~\eqref{eq:potential_energy_1}. We note that the point $x_B$ moves as the simulation proceeds and must be dynamically tracked.  Regarding the value of the attenuation coefficient $\alpha$, we use $\alpha = 0.01$ so as to be able to observe significant attenuation across the computational domain,  which is in order of magnitude comparable to $1/\alpha$. We have verified that using similar values of $\alpha$ modifies the results only marginally; furthermore, we recall that all considered quantities are non-dimensional.



The definition of $\mathcal{U}$ discussed so far in Eq.~(\ref{eq:potential_energy_1}) implies that the SAW acts on all cells, independently of their content: that is, the SAW acts in the same way on all three phases (oil, water, air). This is not realistic, and furthermore it does not account for acoustic radiation pressure.  Namely, in physical experiments, when the SAW reaches a drop, an excess pressure is generated on the free surface of the drop, which leads to a normal stress~\cite{Chu}. To capture this effect we redefine $\mathcal{U}$. One possible option is 
\begin{equation}
    \mathcal{U}(x_{\bm{i}},x_B,\alpha) =\left\{\begin{matrix}
        0 & \begin{tabular}{@{}c@{}}  
            \textrm{liquid cells} \\
            \textrm{detached from solid,}
        \end{tabular}\\ \\
        1 &  \begin{tabular}{@{}c@{}} 
            $x_{\bm{i}}<x_B$  \\  
            \textrm{liquid cells} \\
            \textrm{connected to solid,}
        \end{tabular}\\ \\ 
        e^{-\alpha(x_{\bm{i}}-x_B)} & \begin{tabular}{@{}c@{}} 
            $x_{\bm{i}}>x_B$  \\ 
            \textrm{liquid cells} \\
            \textrm{connected to solid,}
        \end{tabular}\\
\end{matrix}\right.
    \label{eq:potential_energy_2}
\end{equation}
where `detached from solid' means that there is no connection between a considered cell and the solid via other cells filled with liquid (oil or water).  Equation~(\ref{eq:potential_energy_2}) follows 
the physical insight that SAW does not act on air, and therefore there is a jump of $\mathcal{U}$ at the drop free surface. This jump models acoustic radiation pressure through a very simple mechanism.  Consider a liquid (oil or water) cell at the drop surface.  This cell experiences acoustic stress due to the SAW.  If this cell were to detach from the drop (become airborne), it would cease to experience acoustic stress and therefore its energy would decrease. This energy decrease leads to an effective force pushing such a cell out from the surface, and such a force models naturally the acoustic radiation pressure in physical experiments. 

While Eq.~(\ref{eq:potential_energy_2}) includes the desired acoustic radiation pressure effect, we find it convenient to introduce a parameter characterizing its strength, relative to other physical effects.
We are, therefore, led to our final definition

\begin{equation}
    \mathcal{U}(x_{\bm{i}},x_B,\alpha)
    =\left\{
    \begin{array}{ll}
        p & \begin{tabular}{@{}c@{}} 
            $x_{\bm{i}}<x_B$  \\  
            \textrm{liquid cells} \\
            \textrm{detached from solid,}
        \end{tabular}\\ \\
        p\, e^{-\alpha(x_{\bm{i}}-x_B)}  & \begin{tabular}{@{}c@{}} 
            $x_{\bm{i}}>x_B$  \\  
            \textrm{liquid cells} \\
            \textrm{detached from solid,}
        \end{tabular}\\ \\
        1 & \begin{tabular}{@{}c@{}} 
            $x_{\bm{i}}<x_B$  \\  
            \textrm{liquid cells} \\
            \textrm{connected to solid,}
        \end{tabular}\\ \\
        e^{-\alpha(x_{\bm{i}}-x_B)} & \begin{tabular}{@{}c@{}} 
            $x_{\bm{i}}<x_B$  \\  
            \textrm{liquid cells} \\
            \textrm{connected to solid,}
        \end{tabular}\\
    \end{array}
    \right.
    \label{eq:potential_energy_3}
\end{equation}
i.e., we consider that the cells detached from the solid experience an acoustic stress that is a fraction ($p$) of the acoustic stress they would have experienced if they were connected to the solid. If $p=0$ one recovers the more crude approximation given by Eq.~\eqref{eq:potential_energy_2} and if $p=1$ one recovers the acoustic stress that does not account for acoustic radiation pressure given by Eq.~\eqref{eq:potential_energy_1}. The parameter $p$ allows us to control the intensity of the acoustic radiation pressure in our simulations. In the Results section, we present simulations using Eq.~\eqref{eq:potential_energy_3}, in which we account for acoustic radiation pressure; and results using Eq.~\eqref{eq:potential_energy_1}, which does not account for it.

Summarizing, acoustic stress (i.e., the Reynolds stress) in the liquid (oil/water) induces both acoustic flow in the liquid~\cite{Shiokawa1989,shiokawa_dynamics_1994} and acoustic radiation pressure at the free surface~\cite{Chu}. Spatial variations in the acoustic stress along the liquid bulk due to the spatial attenuation of the SAW result in a net bulk force driving flow, reminiscent of the Eckart streaming. Moreover, the smaller acoustic stress acting on those cells disconnected from the substrate leads to  a net force at the free liquid surface. This is reminiscent of acoustic radiation pressure. Both phenomena are different facets of the acoustic stress produced by the SAW.


We do not attribute a significant role to gravity in the extraction process; its primary function is merely to keep the droplet adhered to the substrate. An alternative, more complicated, approach to achieve this end could involve modeling direct interactions between the liquid particles and the substrate. Previous studies, such as that of~\citet{Areshi2019}, have examined such interactions, highlighting their influence on the droplet's equilibrium state, particularly the contact angle. In this work we consider them non-essential for understanding the extraction mechanism, which is primarily governed by SAW effects and the surface tension contrast between the two liquids; hence we omit them. 



To perform the MC simulations, we need to specify the parameter values. The parameters in our model are $J_{oo}$, $J_{ww}$, $J_{ow}$, $g$, $p_{RS}$, $p$, $\alpha$, and implicitly, the system temperature, $T$. While $p_{RS}$, $p$ and $\alpha$ are related to the SAW, and therefore can change in the experimental setup, the $J_{kl}$ are related to the intermolecular forces between the various phases. These three coupling constants $J_{ww}$, $J_{oo}$ and $J_{ow}$ are not independent; since they represent the short-range interaction between the cells, they can be related to the surface tension of the considered liquids.  Therefore, knowing the values of the surface tensions of water and oil allows us to relate the constants $J_{kl}$ to each other. A simple calculation, see the Appendix~\ref{ap:coupling}, leads to 
\begin{equation}
    J_{oo}\approx 0.28 J_{ww} , \hspace{8mm} J_{ow}\approx 0.4 J_{ww} .
\end{equation}
These ratios will be used to set the specific value of the coupling constants $J_{kl}$, as discussed in the next section. 

\section{Results}\label{Results}
\label{sec:results}

\begin{figure*}[t]
\centering
\includegraphics[width=0.9\textwidth]{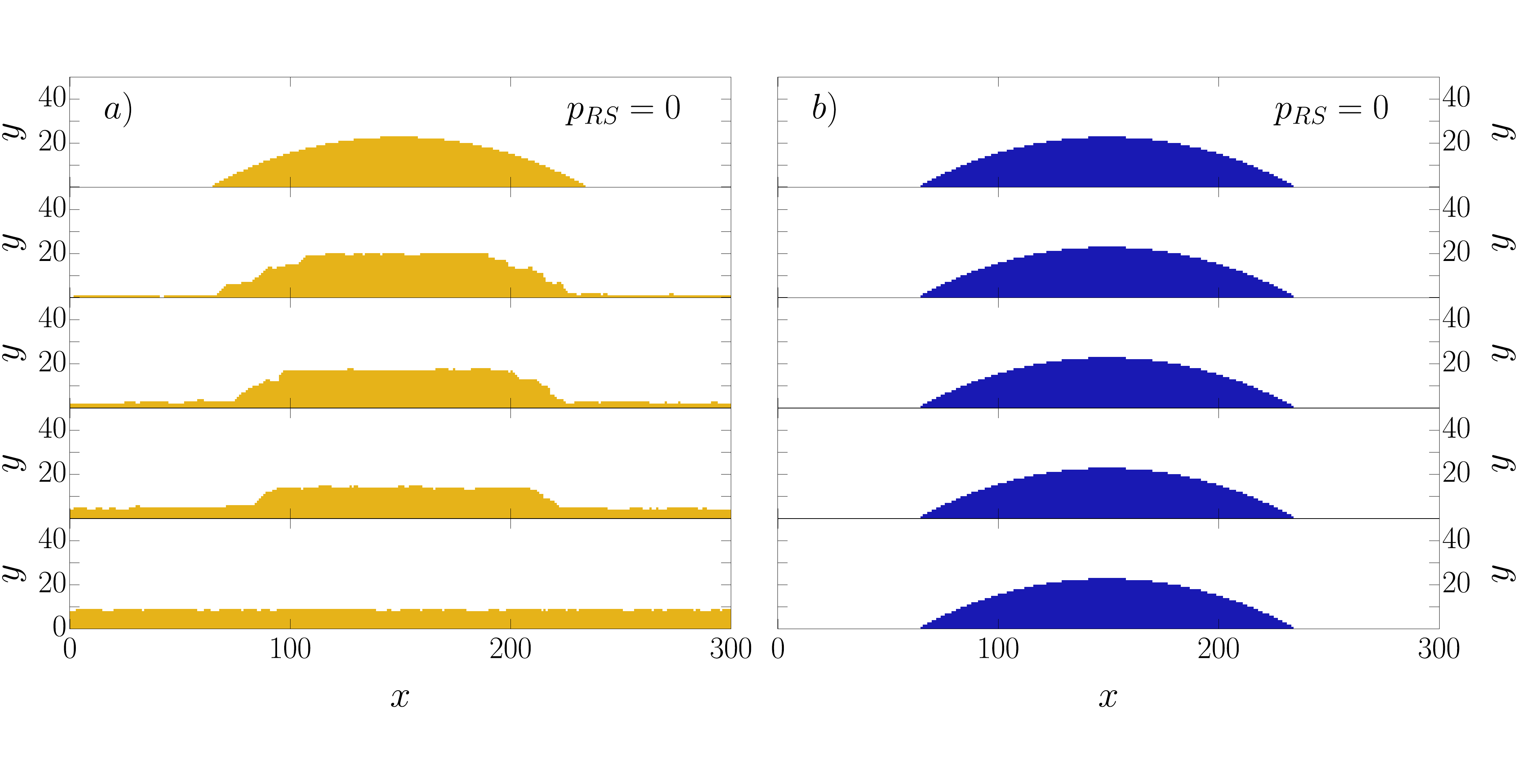}
\caption{Snapshots showing the evolution of a pure system in the absence of SAW ($p_{RS}=0.0$),  (a) oil, and (b) water. Here, $J_{ww}=12.6 $ and $J_{oo}=3.5$. The number of MC steps is $N_{\text{steps}}=10^{11}$. In this and the following figures, $g=20$, the time increases from top to bottom 
at regular intervals, the top figures show the initial condition, water/oil cells are shown in blue/yellow, and vapor cells are not shown.
}
\label{fig:Graf_pure_liquids_NO_SAW}
\end{figure*}

\subsection{Determining interaction energies}

Although we know the ratios of the coupling constants $J_{kl}$, we still have to determine their individual values. To do this we carry out trial simulations in the absence of the SAW for pure liquids and for the emulsion. We know from the experiments~\cite{exp} that, in the absence of the SAW, a drop of pure water maintains its shape, while a drop of pure oil wets the entire substrate. In addition, a drop of emulsion also maintains its shape and, moreover, the oil moves to the free surface of the drop~\cite{exp}. In the absence of the SAW there are only two contributions to the energy present in the system: gravity and the interaction between particles. If gravity is dominant, the droplet spreads out, while if the interaction between particles is dominant, the droplet keeps its initial shape. Therefore, for a given value of $g$, we must choose the $J_{kl}$ such that in the simulations of pure oil, the drop spreads over the entire substrate, in the simulations of pure water the drop maintains the shape, and in the simulations with the emulsion the drop maintains its shape and in addition the oil goes to the surface of the droplet. 
In all simulations that we present, the domain is of size $L_x=300$, $L_y=50$, in the $x$, $y$ directions, and we use $g=20$ and $\rho_w=\rho_o=1$, 
ignoring any buoyancy effects to focus on the acoustic forcing and intermolecular interactions, as noted earlier. The value of $g$ is chosen for convenience; for our purposes, the exact value is relevant only relative to the values of $J_{kl}$. The simulation results are independent of system size; similar results are expected for larger or smaller domains.

The interaction constants $J_{kl}$ are related to the known interfacial energies (surface tensions) of the respective phases. In Appendix~\ref{ap:coupling} we present a simple calculation that allows us to obtain values for the ratios between the constants, $J_{oo}/J_{ww}$ and $J_{ow}/J_{ww}$. All simulations shown respect these ratios, thus, we only need to fix one of the constants to determine all three. We first consider single-phase simulations. Figure~\ref{fig:Graf_pure_liquids_NO_SAW} shows the results for pure systems: only oil (a) and only water (b), in the absence of SAW. For the specified  choices of $J_{ww}$ and $J_{oo}$, we observe that the oil drop wets the entire surface, while the water drop maintains its shape throughout the simulation, as desired. We emphasize that these results are used solely to establish reasonable values for $J_{ww}$ and $J_{oo}$, rather than for the purpose of fully capturing the physics of wetting, which would require incorporating substrate interactions, a crucial factor in studying droplet equilibrium properties \cite{Areshi2019} and spreading behavior \cite{Abraham2002,Lukkarinen1995,Cheng1993}. We also note that extending interactions beyond the first neighbors helps the emulsion droplet retain its shape. In simulations where only interactions with the four nearest neighbors are considered (not shown here), oil particles are more likely to escape from the emulsion.

\begin{figure*}[t]
\centering
\includegraphics[width=0.9\textwidth]{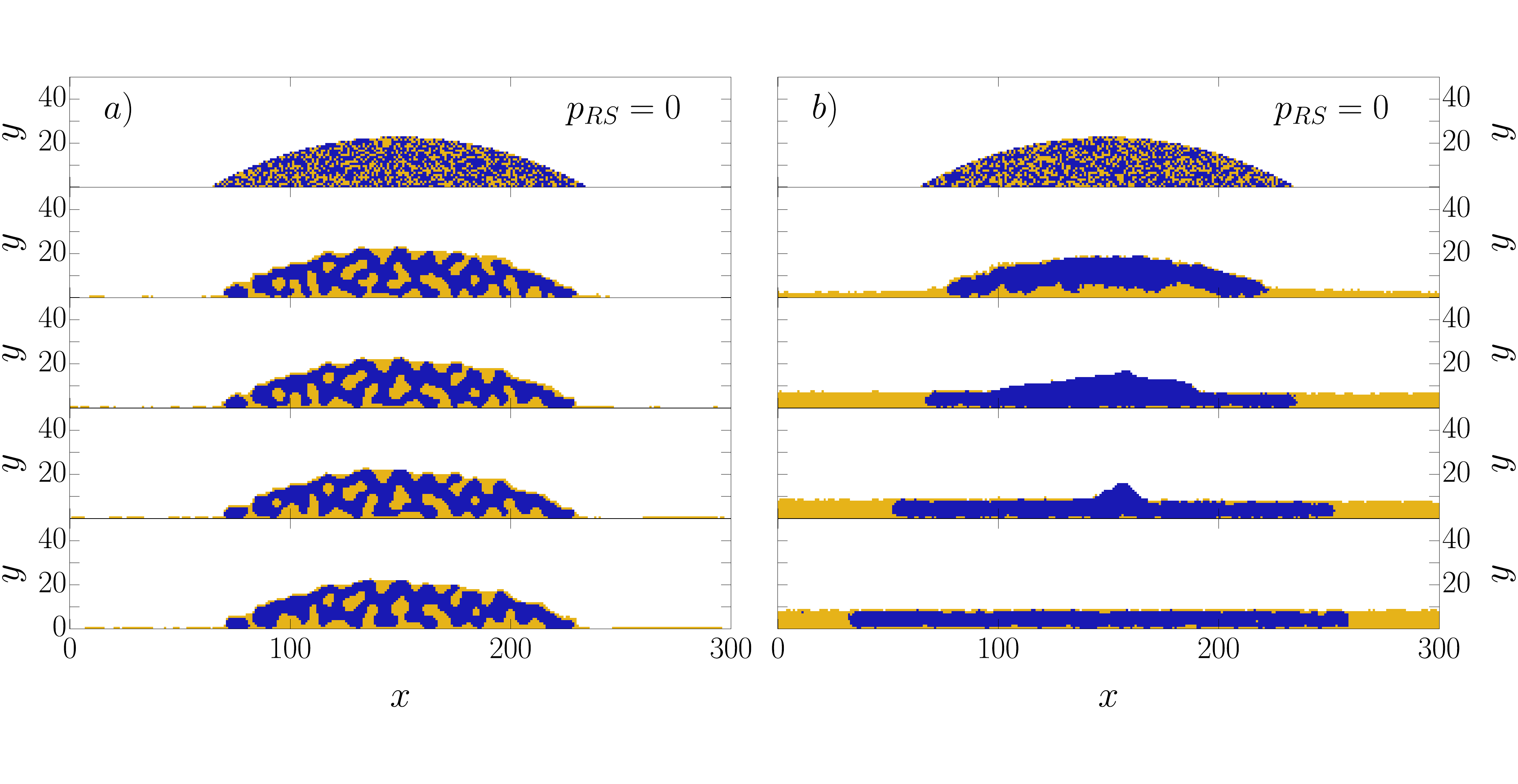}
\caption{Snapshots showing the emulsion evolution in the absence of SAW ($p_{RS}=0.0$) for two different sets of parameters $J_{kl}$: (a)  $J_{ww}=12.6 $, $J_{oo}=3.5$ and $J_{ow}=5.1$,  and (b) $J_{ww}=5.4 $, $J_{oo}=1.5$ and $J_{ow}=2.2$. Here,  $N_{\text{steps}}=10^{11}$ (a) and $N_{\text{steps}}=5\cdot10^{10}$ (b). 
}
\label{fig:Graf_no_SAW}
\end{figure*}

Returning now to the question of appropriate values for the $J_{kl}$, we note that an emulsion droplet must maintain its shape without SAW forcing. Figure~\ref{fig:Graf_no_SAW} shows the results of emulsion drop simulations in the absence of SAW. In (a), we show the results where $J_{kl}$ are large enough for a drop to maintain its shape, while in (b) we show the results where gravity is dominant and the droplet ends up spreading throughout the domain. In order to make our simulations realistic, we choose the coupling constants $J_{kl}$ such that they are large enough to hold the droplet together; in both cases they satisfy the ratios determined in Appendix~\ref{ap:coupling}. We find that the values $J_{ww}=12.6 $, $J_{oo}=3.5$, $J_{ow}=5.1$ (used in Figs.~\ref{fig:Graf_pure_liquids_NO_SAW} and~\ref{fig:Graf_no_SAW}(a)) ensure that desired behavior is obtained, and therefore we use them from here on.  We note in passing the coarsening effect taking place as time progresses, as well as migration of oil particles to the interfaces (both liquid-air and liquid-solid).



\subsection{Influence of SAW on the evolution of emulsions and pure liquids}

With the parameters $J_{kl}$ established, we next carry out simulations including also the SAW. 
Since one of our goals is to identify the contribution of the acoustic radiation pressure to the dynamics of the system in general and to oil extraction in particular, we performed several simulations varying the parameter $p$. 
\begin{figure*}[t]
\centering
\includegraphics[width=0.9\textwidth]{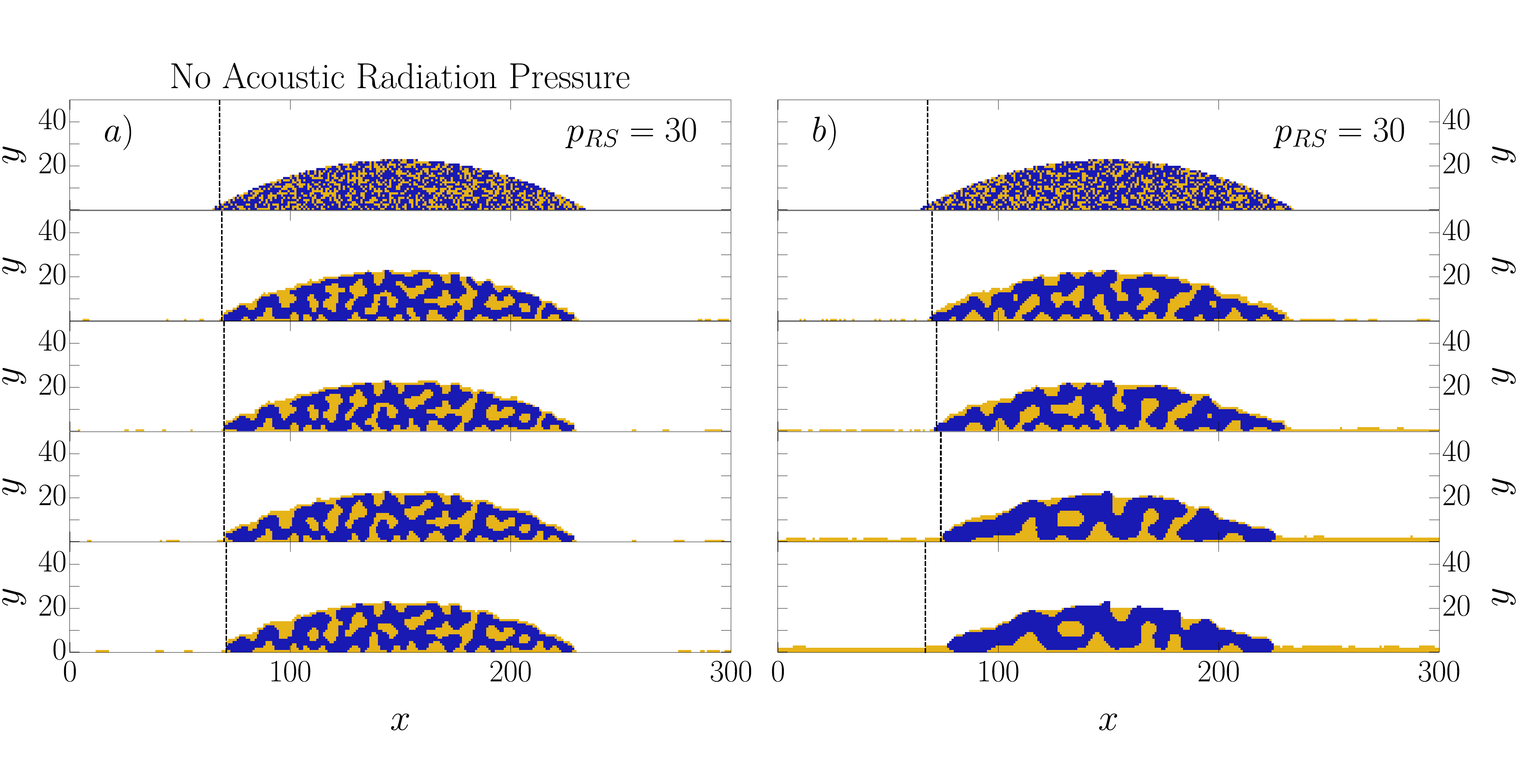}
\caption{Snapshots showing the emulsion evolution in the presence of SAW, where in (a) we account only for acoustic stress in the bulk of the liquid ($p = 1.0$) and in (b) we account for both acoustic stress in the bulk of liquid and the normal stress (acoustic radiation pressure) at the free surface of the drop ($p = 0.9$). The dashed black line represents the position of the effective contact line $x_B$. Note the appearance of thin oil film for late times in (b). The SAW intensity is $p_{RS}=30$, and $N_{\text{steps}}=2\cdot10^{11}$.}
\label{fig:Graf_small_SAW}
\end{figure*}
Figures~\ref{fig:Graf_small_SAW} and~\ref{fig:Graf_high_SAW} show results for weak and strong SAW intensities (characterized by the values of $p_{RS}$), respectively, and demonstrate that in both cases there is a clear difference in the behavior of the system with and without acoustic radiation pressure. In particular, in Fig. \ref{fig:Graf_small_SAW} (weak SAW intensity), it can be seen that the acoustic radiation pressure is a key element that allows for the oil film formation, since in the system without acoustic radiation pressure practically no oil particles escape from the droplet. More specifically, in Fig.~\ref{fig:Graf_small_SAW}(a), where we do not account for acoustic radiation pressure, the drop does not move and there is essentially no influence of SAW. Figure~\ref{fig:Graf_small_SAW}(b), which includes acoustic radiation pressure, shows an oil film appearing on both sides of the drop. Such behavior is similar to the physical experiments~\cite{exp}, although there, the oil film usually is not drawn out in the $-x$ direction (towards the SAW source).   


\begin{figure*}[t]
\centering
\includegraphics[width=0.9\textwidth]{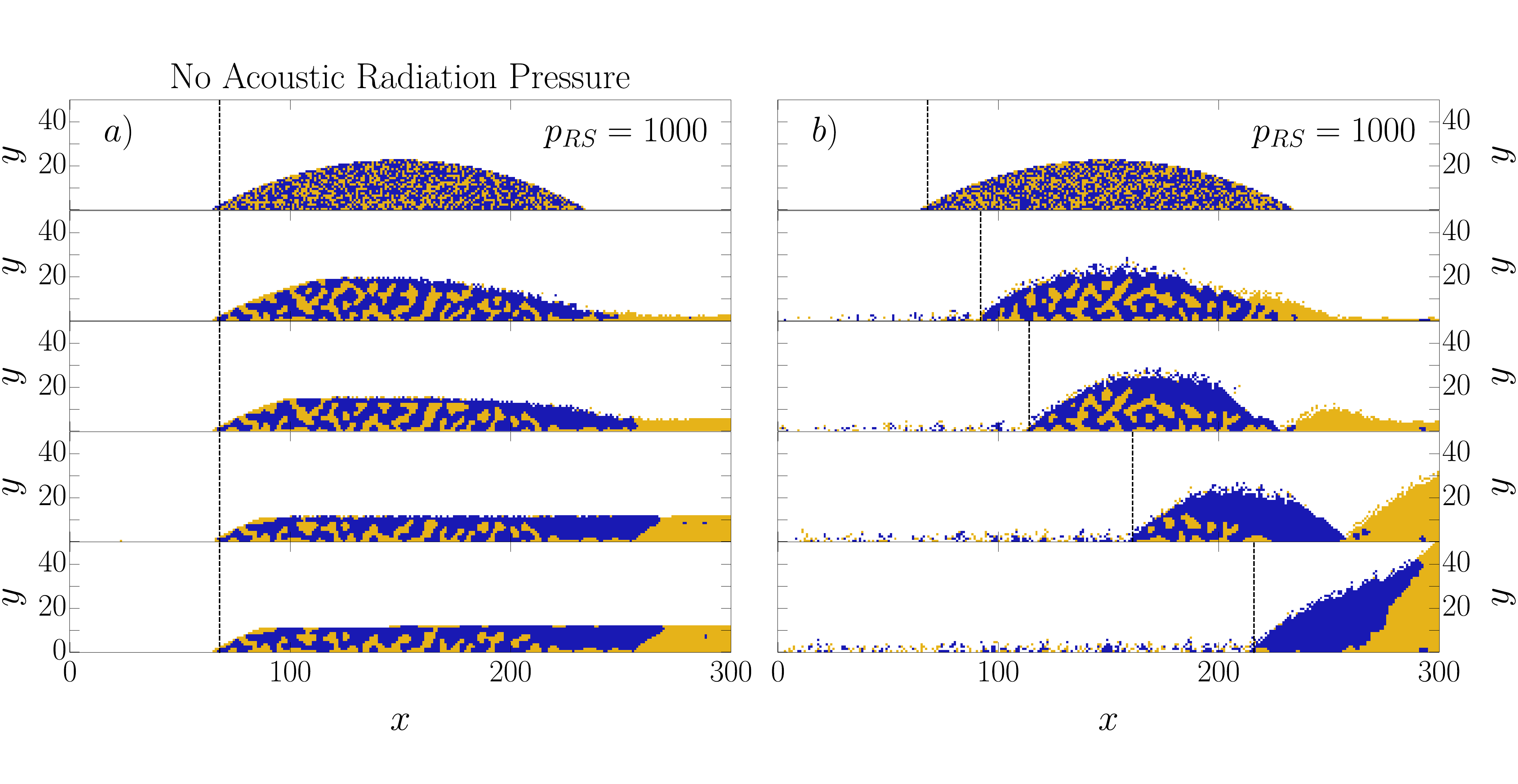}
\caption{The parameters and setup are as in Fig.~\ref{fig:Graf_small_SAW};  $p_{RS} = 1000$, and 
$N_{\text{steps}}=5\cdot10^{8}$.}
\label{fig:Graf_high_SAW}
\end{figure*}

\begin{figure*}[t]
\centering
\includegraphics[width=0.9\textwidth]{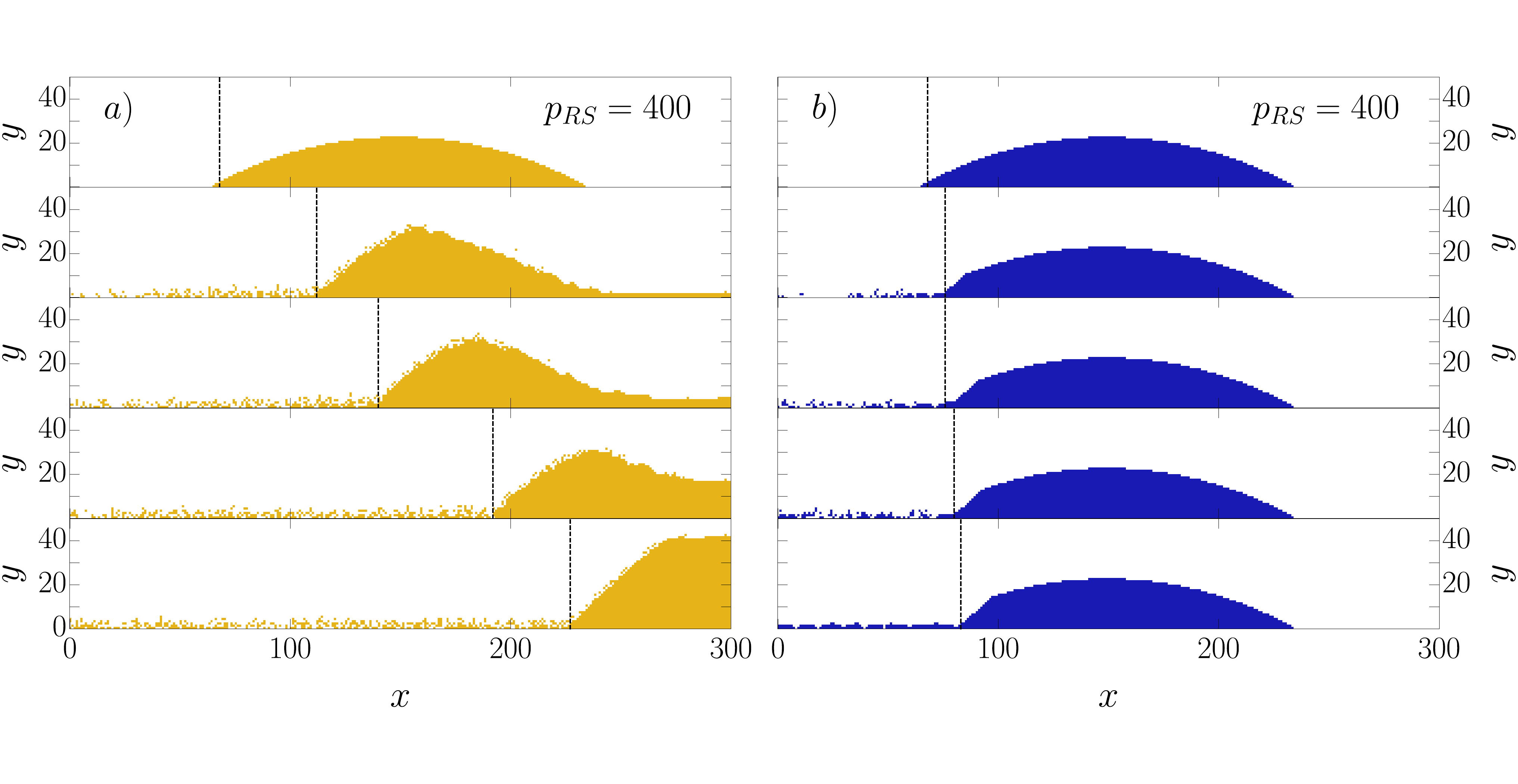}
\caption{Snapshots of the evolution of a pure system in the presence of SAW, for (a) pure oil, and (b) pure water. Here, $p_{RS}=400$, $p=0.9$, $N_{\text{steps}}=10^{9}$.}
\label{fig:Graf_pure_liquids_SAW}
\end{figure*}

Figure \ref{fig:Graf_high_SAW}, where the SAW intensity is much larger, shows more clearly how the acoustic radiation pressure is also a key factor responsible for the dynamics of the simulated droplet. 
Without acoustic radiation pressure, see Fig.~\ref{fig:Graf_high_SAW}(a), we observe that even though the particles are pushed in the SAW propagation direction, the vertical line $x=x_B$, which marks the (left) contact point of the drop, does not move at all. By contrast, once we include acoustic radiation pressure, see Fig.~\ref{fig:Graf_high_SAW}(b), the entire drop is pushed in the SAW propagation direction. For both simulations, careful inspection of the figures shows that the oil moves to the surface of the macroscopic droplet due to the lower interaction energy $J_{oo}$, corresponding to a lower surface tension; and from there the oil is extracted or pushed by the SAW. 

As mentioned in the preceding section, the parameter $p$ allows us to modulate the intensity of the acoustic radiation pressure, since this contribution is maximum when $p=0$ and disappears when $p=1$. For any value $p\in [0,1)$ there is acoustic radiation pressure in the system, and the simulations we have performed exhibit a similar qualitative behavior. In the simulations presented here we choose $p=0.9$, for two reasons: first, when $p$ is zero or very small, the behavior of the system is noisy, and elucidating the effects of the SAW is more difficult; and second, simulations with $p$ very close to unity, while exhibiting behavior qualitatively similar to that observed with $p=0.9$, take longer; therefore, we use $p = 0.9$ as a compromise. 


The SAW has two notable effects on the emulsion drop, occurring in different SAW intensity ranges: the extraction of an oil film; and the streaming of the macroscopic droplet away from the SAW source. For small values of the SAW intensity ($p_{RS}\sim 30$, Fig.~\ref{fig:Graf_small_SAW}) the oil is extracted from the droplet, forming films. For larger values, $p_{RS}\sim 100$, it is also possible to extract water from the droplet, and to have films such that both types of particles are present (results not shown). 
However, only for even larger values, ($p_{RS}\sim 1000$, Fig.~\ref{fig:Graf_high_SAW}) is it possible to see the streaming of a macroscopic emulsion droplet.

For completeness, we also discuss briefly the results obtained by considering pure oil and water drops exposed to SAW. Figure~\ref{fig:Graf_pure_liquids_SAW} shows an example of the results, illustrating that for the same value of $p_{RS}$, the SAW leads to flow of the oil drop, while the equivalent water drop remains stationary, the SAW causing only a minor change in its shape. This difference can only be explained by the differences in the drop interaction energies $J_{oo}$ and $J_{ww}$, i.e., by the differences in their surface tensions. For both oil and water, we observe a small film of particles moving in the opposite direction to the SAW; this behavior is consistent with that found in experiments for pure substances under the action of SAW~\cite{Rezk2012}. 

\subsection{Global time-dependent results illustrating SAW influence}

Finally, we discuss time-dependent results that illustrate the global features of evolving emulsion drops; for convenience we consider sufficiently small values of $p_{RS}$ that the drop remains stationary.  

\begin{figure}[t]
\centering
\includegraphics[width=0.45\textwidth]{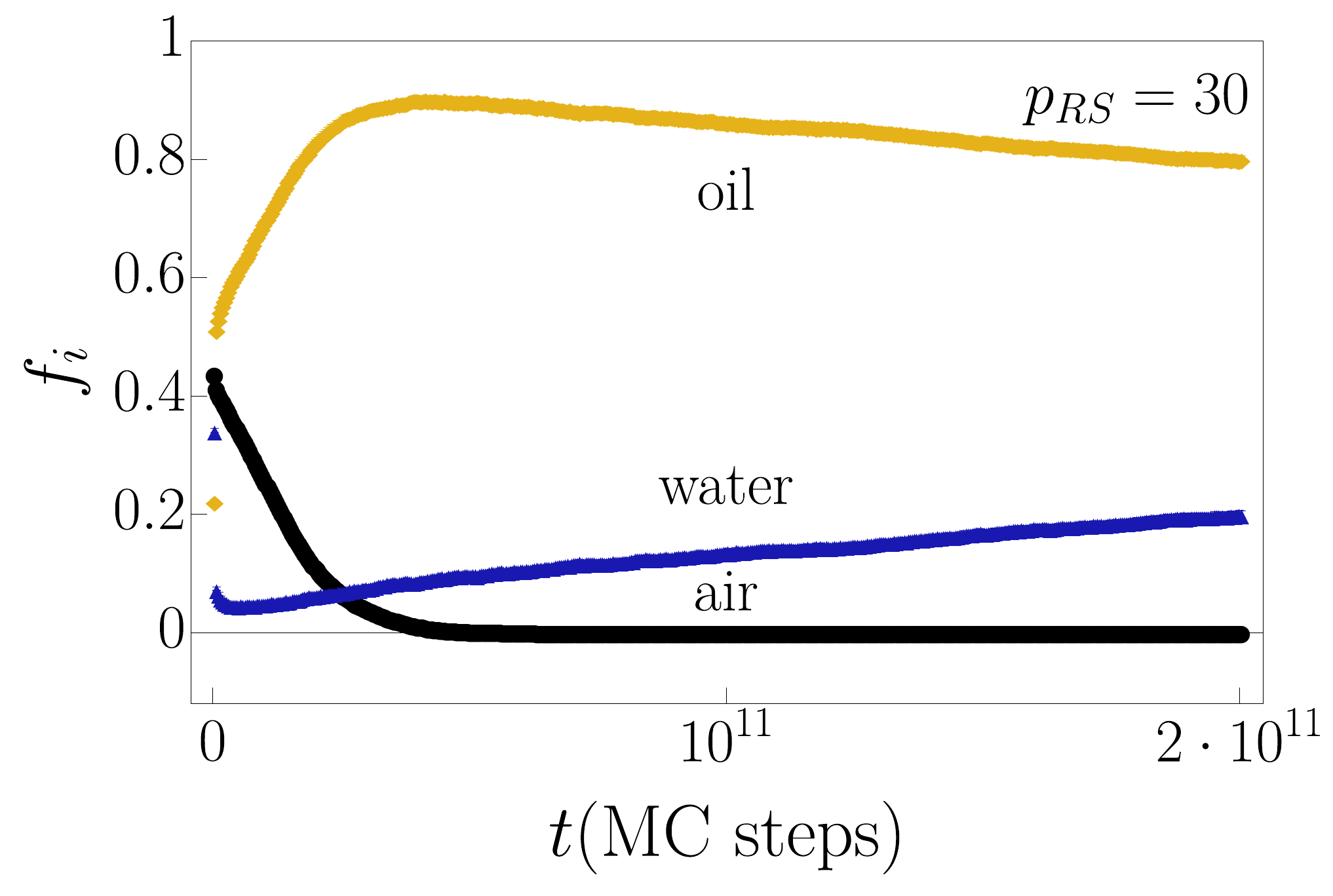}
\caption{Evolution of the fractions $f_i$ (see text for the definition) of oil, water, and vapor/air, averaged over 50 simulation realizations. Water is represented by blue triangles, oil by yellow diamonds, and vapor/air by black circles. Here, $p_{RS}=30$ and $p=0.9$. The error bars are typically smaller than the symbol size.}
\label{fig:surface}
\end{figure}

First, in Figure~\ref{fig:surface}, we track the composition of the system interface over time for $p_{RS} = 30$ and $p=0.9$, where we define the interface to include any ``dry'' areas of the substrate not covered by liquid, as well as the free surface of the bulk liquid.
The figure shows the fractions $f_o, f_w, f_a$ of oil, water and air, respectively, at this interface.
To obtain the results shown in Figure~\ref{fig:surface} we do the following: at every time instant, we loop through the columns of the cell array from top to bottom until a liquid cell, either water or oil, is found. The type of cell we first reach is counted as the cell that is at the surface of the drop or film. If we reach the substrate without finding any liquid cell, we consider that the column is an air one. Due to the initial condition, at $t = 0$ about half of the columns are of the air type, 
and the fraction of water at the free surface of the drop is approximately $1.5$ times larger than the fraction of oil. For early times, since the oil cells are characterized by lower interaction energy, they rapidly move to the surface, replacing the water there (this occurs during the first $10^{10}$ or so MC steps shown in Fig.~\ref{fig:surface}). This is why the fraction of water cells at the surface decays so rapidly, appearing discontinuous on the time scale shown in this figure. 

For intermediate times, the oil fraction at the interface continues to increase and the air fraction starts to decrease as oil begins to be extracted from the droplet and spreads over the dry parts of the surface. Finally, when practically no air columns in the system remain, the behavior of the system changes: the oil fraction starts to (slowly) decrease and the water fraction to increase. This behavior can also be seen in the last snapshot of Fig.~\ref{fig:Graf_small_SAW}(b), which uses the same parameters. This effect occurs since the SAW extracts the oil from the surface of the drop (a process in which the acoustic radiation pressure plays an important role). For longer times, as this oil extraction continues, the amount of oil at the surface of the main droplet decreases and begins to be replaced by water, and there are insufficient available oil cells that can go to the surface to replace the oil being extracted. 

\begin{figure*}[t]
\centering
\includegraphics[width=0.9\textwidth]{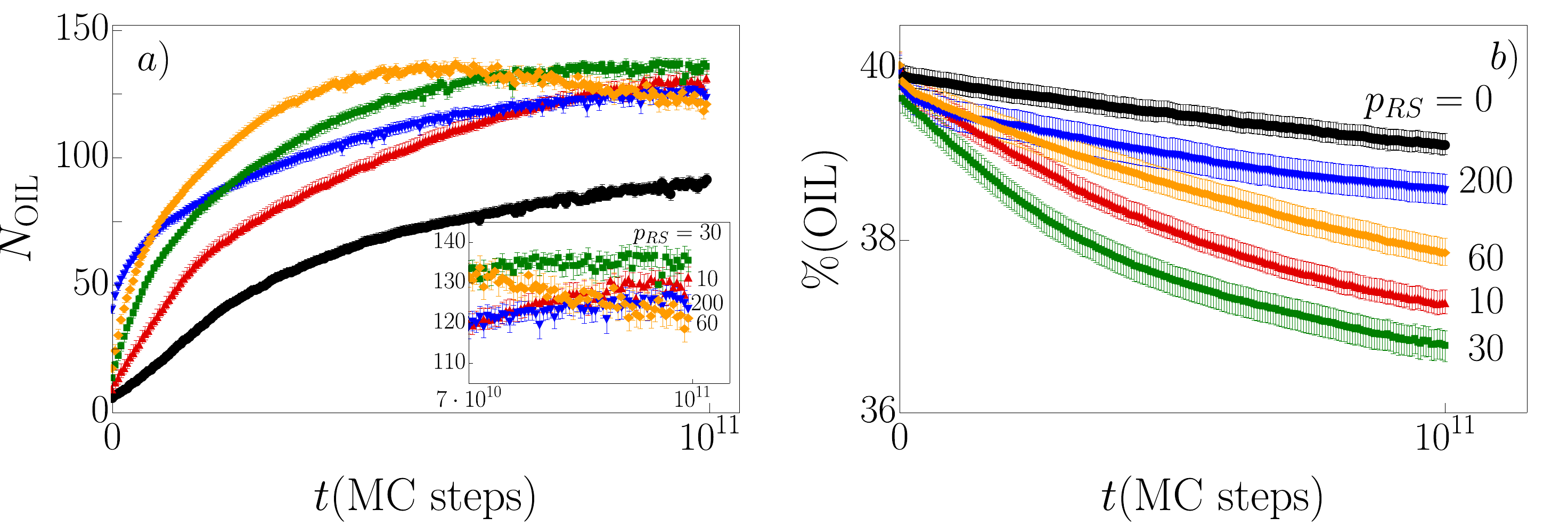}
\caption{(a) Number of oil particles $N_{\mathrm{OIL}}$ left of the original $x_B$ line, i.e. $x_B$ at $t=0$ (see e.g. top snapshots of Fig.~\ref{fig:Graf_small_SAW}), averaged over 50 simulation realization for different values of $p_{RS}$ with $p=0.9$. Inset: zoomed view at later times. (b) Evolution of the drop's oil content (the cells right of the $x_B$ line) averaged over 50 simulation realization for different values of $p_{RS}$ with $p=0.9$. The color scheme is the same in both panels.}
\label{fig:extraction}
\end{figure*}

Figure~\ref{fig:extraction}(a) shows the number of oil particles located to the left of the initial position of the $x_B$ line ($x_B|_{t=0}$). This position can be seen at the top snapshots of Figs.~\ref{fig:Graf_small_SAW}--\ref{fig:Graf_pure_liquids_SAW}. In Fig.~\ref{fig:extraction}(a), we can see that for $p_{RS}=0$ (the black line), quite a few oil particles escape from the droplet, especially at longer times. For short times, the amount of oil escaping from the droplet always increases with $p_{RS}$ ($p=0.9$ in all cases). This is reasonable since, once the oil has accumulated on the surface of the droplet, it becomes easier to extract as the intensity of the SAW increases. However, at longer times, this trend does not hold because, for larger values of $p_{RS}$, such as 200, the system also begins to extract water, which then hinders the extraction of oil from the droplet. This effect does not occur for moderate values of $p_{RS}$, such as 30, where only oil is extracted from the droplet.

Figure~\ref{fig:extraction}(b) shows the time evolution of the oil concentration within the macroscopic droplet, which we define as the group of all cells located to the right of the dynamical (time-dependent) $x_B$ line, that is, the region where the SAW attenuates, for different values of $p_{RS}$ ($p=0.9$ in all cases). At $t=0$, all cases start with a concentration of $c = 0.4$. As time progresses and oil is extracted, this concentration decreases. Even in the absence of SAW excitation ($p_{RS} = 0$), the percentage drops slightly, since some oil particles can still escape from the droplet, as shown in Fig.~\ref{fig:Graf_no_SAW}(a). For small values of $p_{RS}$ its increase leads to a more efficient oil extraction. Nonetheless, beyond a certain SAW intensity, $p_{RS}\sim 50$, the percentage of oil to the right of $x_B(t)$ decreases more slowly than for the smaller intensities. This occurs since for these values the SAW can extract water from the macroscopic droplet as well, and as a consequence, the order of the curves in Fig.~\ref{fig:extraction}(b) is not monotonic in $p_{RS}$.

When analyzing this figure, it is important to keep in mind that, although we measure either the number of oil particles located to the left of the initial $x_B|_{t=0}$ line or the percentage of oil to the right of the (time-dependent) $x_B(t)$ line, both metrics only capture oil extraction in the direction opposite to the SAW, i.e. in the negative direction of the $x$-axis. However, as shown in Fig.~\ref{fig:Graf_small_SAW}(b), a similar oil film forms on the right side of the droplet, in the direction of the SAW source (i.e., in the positive $x$-direction).


The time dependence of the $x_B(t)$ line in Fig.~\ref{fig:extraction}(b) is important to consider when analyzing the results in comparison with Fig.~\ref{fig:extraction}(a), where we measure the number of oil particles located to the left of $x_B|_{t=0}$. For example, the case with $p_{RS} = 200$ shows a slightly higher percentage of oil extracted than the $p_{RS} = 0$ case when measured relative to the time-dependent $x_B$ line. However, in Fig.~\ref{fig:extraction}(a), where we show the number of oil particles located to the left of the original $x_B$ line, the number of oil particles is significantly larger. This effect arises because the $x_B$ line tends to shift leftward at late times, as seen in Fig.~\ref{fig:Graf_small_SAW}(b), meaning that some oil may lie to the left of the original $x_B$ line but to the right of the updated one.

\section{Conclusion}\label{sec12}

In this study, we explore the extraction of oil from an oil-in-water emulsion using a Monte Carlo (MC) based discrete model, incorporating interactions between oil, water, and air, as well as external forces such as gravity and forcing due to surface acoustic waves (SAWs). Our results provide insight into the mechanisms governing oil separation under acoustic excitation and highlight key factors that drive the observed dynamics.

Our simulations confirm that oil naturally migrates toward the surface of the droplet due to its lower surface energy, independent of external forces. When SAW forcing is introduced, it induces both acoustic streaming within the liquid and acoustic radiation pressure at the free surface of the droplet, leading to the formation of an oil film on top of and ahead of the droplet. This process is primarily driven by the difference in surface tensions of oil and water, modeled by appropriate interaction energies.

A critical finding of our study is the essential role of acoustic radiation pressure in enabling oil extraction. In simulations that exclude this effect, oil remains within the droplet and no oil film formation occurs. Conversely, when acoustic radiation pressure is accounted for, an oil film detaches and spreads along the solid substrate, mirroring experimental observations~\cite{exp}. Moreover, as the SAW intensity increases, oil extraction becomes more efficient until a threshold is reached, beyond which water is also displaced. For sufficiently high SAW intensities, not only is the oil extracted, but the entire droplet undergoes flow, illustrating a transition from selective oil removal to bulk fluid motion. 

The main mechanism that allows the oil extraction is the accumulation of oil at the free surface of the drop, which acts as a reservoir from which oil is extracted into the film, under the action of the SAW forcing. While the acoustic stress in the bulk of the drop supports flow along the path of the SAW, the mechanism that forces the separation of an oil film off the drop and the motion of the oil meniscus along the solid substrate is an acoustic-capillary balance between the acoustic stress, which translates to acoustic radiation pressure applied at the free surface of the liquid, and capillary stress. In the case of oil, the acoustic stress dominates in the balance, and the oil phase leaves the drop. In the case of water, the capillary stress dominates, which renders the water phase at rest.  We note that explicit inclusion of liquid/solid interaction forces is not needed for the purpose of explaining oil/water separation.

In conclusion, our study highlights the effectiveness of discrete modeling in capturing the essential physics of oil extraction via SAWs and reinforces the importance of acoustic radiation pressure in this phenomenon. Future work may focus on refining the model to include additional experimental parameters, such as substrate interactions and more complex fluid dynamics, with the goal of further bridging the gap between simulation and real-world applications.

\begin{acknowledgements}

J.M.M acknowledges financial support from Ministerio de Ciencia, Innovación y Universidades (Spain), Agencia Estatal de Investigación (AEI, Spain, 10.13039/501100011033), and European Regional Development Fund (ERDF, A way of making Europe) through Grant No. PID2020-112936GBI00. J.M.M is grateful to the Spanish Ministerio de Universidades for a predoctoral fellowship No. FPU2021-01334. The authors acknowledge the use of the computing facilities of the Instituto de Computación Científica Avanzada of the University of Extremadura (ICCAEx), where the MC simulations were run. Furthermore, 
this work was supported by the BSF grant No. 2020174, and by the donors of ACS Petroleum Research Fund under PRF\# 62062-ND9. J.A.D acknowledges support from Consejo Nacional de Investigaciones Científicas y Técnicas (CONICET, Argentina) with Grant PIP 02114-CO/2021 and Agencia Nacional de Promoción Científica y Tecnológica (ANPCyT, Argentina) with Grant PICT 02119/2020.

\end{acknowledgements}

\appendix
\section{MC simulations details}\label{ap:details}

The simulation performed is a discrete time kinetic Monte Carlo (KMC) simulation with Kawasaki dynamics \cite{Newman1999}, briefly described here. In each simulation step, the algorithm randomly selects two adjacent cells and attempts to swap them. To do this, it computes the energy difference $\Delta \mathcal{H}=\mathcal{H}_{\nu}-\mathcal{H}_{\mu}$ between the final state $\nu$ and the initial state $\mu$ using Eq.~(\ref{eq:energy}). An exchange between the position of cells is accepted or rejected following the Metropolis acceptance criterion, namely \cite{Newman1999}
\begin{equation}
    A(\mu \rightarrow \nu)=\left\{\begin{array}{cc}{e^{- \Delta  \mathcal{H}}} & {\Delta \mathcal{H}>0}\,, \\ {1} & {\Delta  \mathcal{H} \leq 0}\,,\end{array}\right.
    \label{eq:metropolis}
\end{equation}
where $A(\mu \rightarrow \nu)$ is called the acceptance rate for the $\mu \rightarrow \nu$ transition, and $\Delta \mathcal{H}$ is the energy difference between the states (recall that the energy is in units of $k_B T$). The boundary condition is set free, i.e. at the four domain boundaries,  there are walls that the particles cannot cross.

As is customary in KMC simulations, we assume that Eq.~\eqref{eq:metropolis} remains valid even when the system is out of equilibrium. This assumption allows us to use the described algorithm to simulate the evolving system. 

As every simulation realization is different, the uncertainties of the magnitudes presented have been calculated by averaging over all the runs launched for this parameter choice following the jackknife procedure \cite{Young2015,Efron1982}. If $x_i$ is the value of the $x$-quantity at a given time for the $i$-run, then the mean, $\bar{x}$, is defined as: 
\begin{equation}  
  \bar{x} = \frac{1}{N} \sum_{i=1}^N x_i \,,  
\end{equation}  
where $ N $ represents the total number of runs, corresponding to the number of simulations performed.
The $i$-th jackknife estimate of a quantity $ x $ is obtained by averaging over all runs while excluding the data from the $i$-th run:
\begin{equation}
x_i^{\mathrm{JK}}=\frac{1}{N-1}\sum_{k=1, k \neq i}^N x_k\,.
\end{equation}
The variance of $\bar{x}$ is then defined as
\begin{equation}
    \sigma_\mathrm{JK}(\bar{x})=\frac{N-1}{N} \sum_{k=1}^N(\bar{x}-x_i^\mathrm{JK})^2\,. 
    \label{eq3:sigmaJK}
\end{equation}
Thus, the estimated value is given by $ \bar{x} \pm \sqrt{\sigma_\mathrm{JK}} $ (within one standard deviation).

\section{Relation between the coupling constants $J_{kl}$ and the surface tensions of the liquids}\label{ap:coupling}

In this section we present a simple calculation that relates the coupling constants of the particles in our model to the macroscopic surface tensions of water, $\gamma_w$, oil, $\gamma_o$, and the interfacial surface tension of oil and water, $\gamma_{ow}$, whose values are known. 

If $J_{ww}$ is the typical binding energy for the interaction between two water particles, then the binding energy per particle of water in the bulk is 
\begin{equation}
    E_{w,b}=-\frac{1}{2}J_{ww}Z_b,
\end{equation}
where $Z_b$ is the average number of neighbors and the factor 1/2 appears to avoid double-counting interactions. Similarly, for the binding energy per particle of water in the surface we have
\begin{equation}
    E_{w,s}=-\frac{1}{2}J_{ww}Z_s,
\end{equation}
where $Z_s$ is the average number of neighbors for a particle of water in the surface. As the surface tension is the energy required to create an interface per unit area, then
\begin{equation}
    \gamma_w=\frac{1}{2a}J_{ww}\left(Z_b-Z_s\right),
    \label{eq:surface_tension_water}
\end{equation}
where $a$ is the typical area occupied by a particle in the surface, and surface tension is in units of $k_B T/a$.  Similarly, for oil we have
\begin{equation}
    \gamma_o=\frac{1}{2a}J_{oo}\left(Z_b-Z_s\right),
    \label{eq:surface_tension_oil}
\end{equation}
where the parameters $a$, $Z_s$ and $Z_b$ are, generally, not the same for oil and water. However, if we assume that they are similar, we find
\begin{equation}
    \frac{\gamma_w}{\gamma_o}\approx \frac{J_{ww}}{J_{oo}}.
    \label{eq:surface_tension_1}
\end{equation}
The interfacial tension $\gamma_{ow}$ between oil and water can be computed as
\begin{equation}
    \gamma_{ow}+\Delta W_{ow}=\gamma_w+\gamma_o,
\end{equation}
 where $\Delta W_{ow}$ is the work per unit area needed to separate an oil-water interface into two interfaces (water-air and oil-air). This work can be estimated as
\begin{equation}
    \Delta W_{ow}=\frac{Z_b-Z_s}{a}J_{ow},
\end{equation}
where we assume that the number of pairs per unit area of the oil-water interface is the same as at the free surface of water and oil, i.e., $Z_s$ and $Z_b$ are the same as in Eqs.\ \eqref{eq:surface_tension_water} and \eqref{eq:surface_tension_oil}. This leads to
\begin{equation}
    \gamma_{ow}=\frac{Z_b-Z_s}{a}\left[\frac{1}{2}\left(J_{ww}+J_{oo}\right)-J_{ow}\right].
\end{equation}
From here it can be easily seen that
\begin{equation}
    \frac{\gamma_{ow}}{\gamma_{w}}=1+\frac{J_{oo}}{J_{ww}}-2\frac{J_{ow}}{J_{ww}},
\end{equation}
and
\begin{equation}
    J_{ow}=\frac{J_{ww}}{2}\left[1+\frac{J_{oo}}{J_{ww}}-\frac{\gamma_{ow}}{\gamma_{w}}\right]\approx\frac{J_{ww}}{2}\left[1+\frac{\gamma_{o}}{\gamma_{w}}-\frac{\gamma_{ow}}{\gamma_{w}}\right].
    \label{eq:surface_tension_2}
\end{equation}
If we take into account the experimental values of  $\gamma_w$, $\gamma_o$ and $\gamma_{ow}$~\cite{Peters2013} 
$\gamma_o/\gamma_w\approx 0.28$ and $\gamma_{ow}/\gamma_w\approx0.5$, then the values of the $J_{oo}$ and $J_{ow}$ in terms of $J_{ww}$ that follow from Eq.\ \eqref{eq:surface_tension_1} and \eqref{eq:surface_tension_2} are
\begin{equation}
    J_{oo}\approx 0.28 J_{ww} ,\hspace{8mm} J_{ow}\approx 0.4 J_{ww}.
    \label{eq:estimate}
\end{equation}
The approximation that $a$, $Z_s$ and $Z_b$ are the same at the three interfaces—water-air, oil-air, and water-oil— may be rather crude. Nonetheless, we expect $J_{ww}$ to be much larger than $J_{oo}$ as the water-water interaction (hydrogen bonding) is stronger than the oil-oil interaction (van der Waals), so the estimates given by Eq.~\eqref{eq:estimate} are reasonable for our purposes.

\end{document}